\documentclass[pra,twocolumn,aps,superscriptaddress,showpacs,floatfix]{revtex4}
\usepackage{amsmath, amssymb, psfrag, tabls, dcolumn, bm, color}
\usepackage[sort&compress]{natbib}	% citation style (1,2,3-->1-3)
\usepackage[pdftex]{graphicx}		% use pdflatex
\usepackage{color}
\usepackage{amsmath}
\usepackage{enumerate}

\newcommand{\be}{\begin{equation}}
\newcommand{\beq}{\begin{equation}}
\newcommand{\ee}{\end{equation}}
\newcommand{\bea}{\begin{eqnarray}}
\newcommand{\eea}{\end{eqnarray}}
\newcommand{\ba}{\begin{array}}
\newcommand{\ea}{\end{array}}

\renewcommand{\vr} {{\bf r}}

\newcommand{\vR} {{\bf R}}

\begin{document}

\title{Colle-Salvetti-type local density functional for the \\
exchange-correlation energy in two dimensions}

\author{S. {\c S}akiro{\u g}lu}
\affiliation{Nanoscience Center, Department of Physics, University of Jyv\"askyl\"a, P.O. Box 35, FI-40014 Jyv\"askyl\"a, Finland}
\affiliation{Physics Department, Faculty of Arts and Sciences, Dokuz Eyl{\"u}l University, 35160 \.Izmir, Turkey}

\author{E. R{\"a}s{\"a}nen}
\email{erasanen@jyu.fi}
\affiliation{Nanoscience Center, Department of Physics, University of Jyv\"askyl\"a, P.O. Box 35, FI-40014 Jyv\"askyl\"a, Finland}

\begin{abstract}
We derive an approximate local density functional for the exchange-correlation 
energy to be used in density-functional calculations of two-dimensional systems. 
In the derivation we employ the Colle-Salvetti wave function within the scheme
of Salvetti and Montagnani [Phys. Rev. A \textbf{63}, 052109 (2001)] to satisfy
the sum rule for the exchange-correlation hole. We apply the functional for the
two-dimensional homogeneous electron gas as well as to a set of quantum dots 
and find a very good agreement with exact reference data.
\end{abstract}


\date{\today}

\maketitle

%\section {Introduction}

Development in modern technology has enabled the fabrication of nanoscale 
electronic devices with a large variety of low-dimensional systems.
Two-dimensional (2D) quantum dots (QDs) are particularly interesting 
examples due to the tunability in their size and shape, and number
of confined electrons~\cite{kou,rei}. 
From the theoretical point of view, QDs constitute an ideal platform
to study the many-particle problem, electronic correlations, and the 
role of the dimensionality.

In density-functional theory~\cite{dft} (DFT) particle-particle interactions 
beyond the classical (Hartree) term are captured through the 
exchange-correlation (xc) functional, which is approximated in practice.
The development of xc functionals of varying portions of
simplicity and accuracy have a long and successful history~\cite{functionals}.
The Colle-Salvetti (CS) scheme~\cite{ColleSalvetti_75,ColleSalvetti_79} 
and its variants~\cite{moscardoreview} 
have had an important role in the development, especially in terms of 
the electronic correlation. However, these efforts have focused
almost solely on three dimensions (3D), apart from orbital functionals
where the aspect of dimensionality is inbuilt through the Kohn-Sham
orbitals. Only very recently, several local~\cite{cs2d,simple} and 
semi-local~\cite{x,xring,gga,gamma,c,c2} 
functionals have been developed in 2D, and in many test cases involving, e.g., 
different QDs, they have outperformed the commonly used 2D local-density 
approximation based on the exact exchange and correlation of the 
homogeneous 2D electron gas~\cite{rajagopal,attaccalite} (2DEG).

In Ref.~\cite{cs2d} a 2D local density functional for the 
correlation energy was derived using the CS framework with a Gaussian
summation for the pair density~\cite{moscardo}. Despite the good
performance of this functional for the correlation, a compatible 
approximation for the exchange energy is needed in view of, e.g.,
total-energy calculations. In fact, a combination with 
the 2D generalized-gradient approximation for the exchange~\cite{gga}, 
leads to a reasonable accuracy in the total energy~\cite{ijqc}. 
However, this combined functional
is still semi-local, i.e., it depends on the density gradients, 
which reduces the numerical efficiency.

In this work we employ the CS framework to derive a 2D {\em local} 
functional for the xc energy, so that both the exchange and correlation are 
treated in the same footing. 
In the derivation we follow the 3D 
scheme of Salvetti and Montagnani~\cite{salmon}
for the second-order
density matrix to satisfy the sum rule of the 
xc hole, which is used to obtain a local density functional of a 
simple polynomial form.
We optimize two remaining parameters of the functional by
fitting against exact results for six-electron QDs. The obtained
parameters show universality in the sense that
a good accuracy and consistency is found when testing the functional for 
QDs with a varying electron number as well as for the 2DEG.

%In this paper we extend the appoximation for the exchange-correlation functional introduced by Salvetti and Montagnani in Ref.~\cite{Salvetti_01} to the two-dimensional case.
%The approach is based on revising the widely used~\cite{Moscardo_91,Singh_99,Tao_01,Imamura_02,Pan_06} Colle-Salvetti (CS)
%correlation functional~\cite{ColleSalvetti_75,ColleSalvetti_79}, which is  based on the Hartree-Fock second-order density matrix, by forcing it to
%satisfy the sum rule of the exchange-correlation hole. Despite it is inspired from CS scheme, there are deep differences between them.~\cite{Moscardo_06}
%Applications to several few-electron QDs with various relative amounts of correlation confirm the promising performance of the resulting functional.

%\section{Derivation of the functional}

The electron-electron interaction energy can be {\it formally} expressed 
(in Hartree atomic units) as
\begin{equation}
E_{ee}=\langle\Psi|\hat{V}_{ee}|\Psi\rangle=\int d\mathbf{r}_{1}\int d\mathbf{r}_{2}\frac{\rho_{2}(\mathbf{r}_{1},\mathbf{r}_{2})}{|\mathbf{r}_{1}-\mathbf{r}_{2}|},
\label{eee}
\end{equation}
where
\begin{eqnarray}
\rho_2(\vr_1,\vr_2) &=& \frac{N(N-1) }{2}
\sum_{\sigma_1,\sigma_2} \int{d 3}...\int dN \nonumber \\
&\times& |\Psi(\vr_1 \sigma_1,\vr_2 \sigma_2,3, ...,N)|^2.
\end{eqnarray}
is the diagonal element of the spinless second-order density matrix
describing the distribution density of electron pairs.
Here $\int dN$ denotes the spatial integration and spin summation
over the $N$th spatial and spin coordinates $\vr_N\sigma_N$, and
$\Psi(1,2,\ldots,N)$ is the ground-state many-body wave function.
The element $\rho_2(\vr_1,\vr_2)$ satisfies the
normalization 
\begin{equation}
\frac{N(N-1)}{2}=\int d\mathbf{r}_{1} \int d\mathbf{r}_{2} \, \rho_{2}(\mathbf{r}_{1},\mathbf{r}_{2}),
\end{equation}
and it is related with the electron density, i.e., 
the diagonal term of the first-order density matrix through
\begin{equation}\label{FSDM}
\rho(\mathbf{r}_{1})=\frac{2}{N-1}\int d\mathbf{r}_{2}\,\rho_{2}(\mathbf{r}_{1},\mathbf{r}_{2}).
\end{equation}

Next, introducing a symmetric function accounting for all nonclassical effects called 
the {\it pair correlation function} $h(\mathbf{r}_{1},\mathbf{r}_{2})$ 
suggests us to write~\cite{ParrYang_book}
\begin{equation}
\rho_{2}(\mathbf{r}_{1},\mathbf{r}_{2})=\frac{1}{2}\rho(\mathbf{r}_{1})\rho(\mathbf{r}_{2})[1+h(\mathbf{r}_{1},\mathbf{r}_{2})].
\end{equation}
The important sum rule for the xc hole can be expressed in
terms of the pair correlation function as
\begin{equation}\label{sumrule}
\int d\mathbf{r}_{2}\,\rho(\mathbf{r}_{2})h(\mathbf{r}_{1},\mathbf{r}_{2})=-1.
\end{equation}
The total interaction energy in Eq.~(\ref{eee})
can be split in the classical Coulomb (Hartree) term associated with $\rho(\mathbf{r})$ and the
nonclassical (indirect) part associated with the xc energy,
\begin{eqnarray}\label{Vee}
E_{ee} & = & E_{H}+E_{xc} =\frac{1}{2}\int d\mathbf{r}_{1} d\mathbf{r}_{2}\frac{\rho(\mathbf{r}_{1})\rho(\mathbf{r}_{2})}{|\mathbf{r}_{1}-\mathbf{r}_{2}|}\nonumber \\
& + & \frac{1}{2}\int d\mathbf{r}_{1} d\mathbf{r}_{2}\frac{\rho(\mathbf{r}_{1})\rho(\mathbf{r}_{2})h(\mathbf{r}_{1},\mathbf{r}_{2})}{|\mathbf{r}_{1}-\mathbf{r}_{2}|}.
\end{eqnarray}
It should be noted that, compared with the conventional DFT formalism, $E_{xc}$ as 
defined here neglects the kinetic-energy contribution to the
correlation energy. 
The Hartree term can be computed in a straightforward fashion, but the integration
in the xc part is nontrivial due to the pair correlation function $h(\mathbf{r}_{1},\mathbf{r}_{2})$.
The key point in the present work is to obtain an approximation for
$h(\mathbf{r}_{1},\mathbf{r}_{2})$ satisfying the sum rule in Eq.~(\ref{sumrule}).
Before proceeding with that, we will briefly introduce the CS approach which
is relevant for the derivation.

The CS scheme starts with the following ansatz for the many-body 
wave function~\cite{ColleSalvetti_75,ColleSalvetti_79}
\begin{equation}\label{wf}
\Psi(\mathbf{r}_{1}\sigma_{1},...,\mathbf{r}_{N}\sigma_{N})=\Psi_{\rm HF}(\mathbf{r}_{1}\sigma_{1},...,\mathbf{r}_{N}\sigma_{N})
\prod\limits_{i>j}[1-\varphi(\vr_{i},\vr_{j})],
\end{equation}
where HF refers to the single-determinant Hartree-Fock wave function, and
\begin{equation}\label{cwf}
\varphi(\vr_{1},\vr_{2})=[1-\Phi(\vR)(1+\zeta\,r)]\,\exp{[-\beta^{2}(\vR)\,r^2]}
\end{equation}
describes the correlated part of the wave function written in 
center-of-mass, $\vR=(\vr_{1}+\vr_{2})/2$, and relative,
$\mathbf{r}=\vr_{1}-\vr_{2}$ coordinates. 
The parameter $\zeta$ comes from the cusp conditions, and the quantities $\Phi$ and $\beta$ 
act as correlation factors. In Refs.~\cite{moscardo} and ~\cite{cs2d}
dealing with 3D and 2D systems, respectively, $\beta$ was introduced as a {\em local} factor for 
the correlation length, $\beta(\vR)=q\,\rho^{1/D}(\vR)$, where $D$ is the dimension, 
$q$ is a fitting parameter, and $\rho(\vR)$ is the electron density. The CS approach assumes that 
the first and second-order density matrices can be written as
$\rho_{1}(\vr_{1},\vr_{2})=\rho_{1}^{\rm HF}(\vr_{1},\vr_{2})$ and 
$\rho_{2}^{\rm CS}(\vr_{1},\vr_{2})=\rho_{2}^{\rm HF}(\vr_{1},\vr_{2})\,[1-\varphi(\vr_{1},\vr_{2})]^{2}$,
respectively~\cite{Tao_01}.

To approximate $h(\mathbf{r}_{1},\mathbf{r}_{2})$, we extend the strategy by
Salvetti and Montagnani~\cite{salmon} to 2D by introducing the 
correlation factors in the following way:
\be
\beta(\mathbf{r}_{1},\mathbf{r}_{2})=\gamma\,\rho^{1/2}(\mathbf{r}_{1})\rho^{1/2}(\mathbf{r}_{2}),
\label{a}
\ee
\be
\Phi=\frac{\beta^{\alpha}}{\sqrt{\pi}+\beta^{\alpha}},
\label{b}
\ee
\be
\varphi(\beta)=\left[1-\Phi(1+r)\right]\Phi\,e^{-\beta r^{2}} 
\label{c}
\ee
with $r=|\mathbf{r}_{1}-\mathbf{r}_{2}|$.
Above, $\gamma$ and $\alpha$ are optimizable parameters ($\gamma$ with dimension of $\rho^{-1}$), and
$\Phi$ is a monotonic function varying between zero and one. The differences
from the original CS scheme are obvious; most importantly, $\beta$ is
now a non-local functional of the density.

We may now search for the pair correlation function
\begin{equation}\label{PCF}
h(\mathbf{r}_{1},\mathbf{r}_{2})=\frac{\varphi^{2}-2\,\varphi}{f},
\end{equation}
where $f$ is assumed to be a simple polynomial of the form
\begin{equation}
f(\Phi)=a_{0}\Phi^{n}+a_{1}\Phi^{n-1}+...
\end{equation}
The nominator in the expression for $h(\mathbf{r}_{1},\mathbf{r}_{2})$ is similar to the CS 
functional~\cite{ColleSalvetti_75, ColleSalvetti_79},
whereas the denominator is chosen such that to the sum rule in Eq.~(\ref{sumrule}) is satisfied.
Substituting Eq.~(\ref{PCF}) into Eq.~(\ref{sumrule}) yields
\begin{multline}
\int d\mathbf{r}_{2}\,\rho(\mathbf{r}_{2})h(\mathbf{r}_{1},\mathbf{r}_{2})=
\int d\mathbf{r}\rho(\mathbf{r}_{1}+\mathbf{r})h(\mathbf{r}_{1},\mathbf{r}_{1}+\mathbf{r})\\
=\int d\mathbf{r}\frac{\rho(\mathbf{r}_{1}+\mathbf{r})}{f}\Bigg\{\Phi^{4}e^{-2\beta r^{2}}(1+r)^{2}\\
-2\Phi^{3}e^{-2\beta r^{2}}(1+r)+\Phi^{2}e^{-\beta r^{2}}\\
\times\left[e^{-\beta r^{2}}+2(1+r)\right]-2\Phi e^{-\beta r^{2}} \Bigg\}=-1
\end{multline}
This expression involves integrals which can be computed by 
using mean value theorem and the regularity of the functions. 
By following the procedure of Ref.~\cite{salmon},
we obtain
\begin{multline}
\int d\mathbf{r}\,g(\mathbf{r}_{1},\mathbf{r}_{1}+\mathbf{r})\,e^{-b(\mathbf{r}_{1},\mathbf{r}_{1}+\mathbf{r})r^{2}}\,r^{n}\\
\simeq 2\pi g(\mathbf{r}_{1})\int dr e^{-br^{2}}\,r^{n+1}.
\end{multline}
Utilizing this approximate integration, which becomes more accurate as $b$ becomes large, 
leads to
\begin{multline}
\int d\mathbf{r}_{2}\rho(\mathbf{r}_{2})h(\mathbf{r}_{1},\mathbf{r}_{2}) \\ 
\simeq \frac{2\pi\rho(\mathbf{r}_{1})}{f\,\beta}\Big[\Phi^{4}(i_{0}+2i_{1}+i_{2})-\Phi^{3}(2i_{0}+2i_{1})\\
+\Phi^{2}(i_{0}+2j_{0}+2j_{1})-2\Phi j_{0}\Big]=-1,
\end{multline}
where we define $i_{n}$ and $j_{n}$ as
\begin{equation}
i_{n}=\sqrt{2^{-n-2}\beta^{-n}}\int dx~e^{-x^{2}}x^{n+1}
\end{equation}
and
\begin{equation}
j_{n}=\sqrt{\beta^{-n}}\int dx~e^{-x^{2}}x^{n+1}.
\end{equation}

Using the definition of $\beta$ and calculating the integrals leads to the final result for the polynomial function,
\begin{equation}\label{Ffunction}
f=-2\left(\frac{\pi}{\gamma}\right)\Phi(a_{0}\Phi^{3}+a_{1}\Phi^{2}+a_{2}\Phi-1),
\end{equation}
where the coefficients are given by
\begin{equation}
\label{coeff_a}
\begin{array}{c}
\displaystyle a_{0}=\frac{1}{4}+\frac{1}{8\beta}+\frac{1}{4}\left(\frac{\pi}{2\beta}\right)^{1/2},\\
\displaystyle a_{1}=-\frac{1}{2}-\frac{1}{4}\left(\frac{\pi}{2\beta}\right)^{1/2},\\
\displaystyle a_{2}=\frac{5}{4}+\frac{1}{2}\left(\frac{\pi}{\beta}\right)^{1/2}.\\
\end{array}
\end{equation}

Computation of the integral in Eq.~(\ref{Vee}) is performed by a similar procedure,
\begin{eqnarray}
E_{xc} & = & \frac{1}{2}\int d\mathbf{r}_{1} d\mathbf{r}_{2}\frac{\rho(\mathbf{r}_{1})\rho(\mathbf{r}_{2})h(\mathbf{r}_{1},\mathbf{r}_{2})}{|\mathbf{r}_{1}-\mathbf{r}_{2}|}\nonumber\\
&=& \displaystyle \frac{1}{2}\int d\mathbf{r}_{1}\rho(\mathbf{r}_{1})\int d\mathbf{r}\rho(\mathbf{r}_{1}+\mathbf{r})\frac{\varphi^{2}-2\varphi}{f\,r}\\
& = & \displaystyle \pi\int d\mathbf{r}_{1}\rho^{2}(\mathbf{r}_{1})\frac{\Phi}{f}\sqrt{\frac{\pi}{\beta}}
\left[b_{0}\Phi^{3}+b_{1}\Phi^{2}+b_{2}\Phi-1\right],\nonumber
\end{eqnarray}
where the coefficients are given by
\begin{equation}
\label{coeff_b}
\begin{array}{c}
\displaystyle b_{0}=\frac{1}{2\sqrt{\pi\beta}}+\frac{1}{2\sqrt{2}}+\frac{1}{8\beta\sqrt{2}},\\
\displaystyle b_{1}=-\frac{1}{2\sqrt{\pi\beta}}-\frac{1}{\sqrt{2}},\\
\displaystyle b_{2}=\frac{1}{\sqrt{\pi\beta}}+\frac{1}{2\sqrt{2}}+1.\\
\end{array}
\end{equation}

Using Eq.~(\ref{Ffunction}) and definition of $\beta$ we obtain the final result for the xc energy in 2D,
\begin{equation}
\label{xc_energy}
E_{xc}=\int d\mathbf{r}\rho^{3/2}(\mathbf{r})\,q(\rho)
\end{equation}
with
\begin{equation}
\label{q}
q(\rho)=-\sqrt{\frac{\pi\gamma}{4}}\left(\frac{b_{0}\Phi^{3}+b_{1}\Phi^{2}+b_{2}\Phi-1}{a_{0}\Phi^{3}+a_{1}\Phi^{2}+a_{2}\Phi-1}\right).
\end{equation}

%\section {Determining the parameters}

The remaining task is to find a reasonable pair of values for $\gamma$ and $\alpha$ 
which determine $q(\rho)$ through Eqs.~(\ref{a}), (\ref{b}), 
(\ref{coeff_a}), and (\ref{coeff_b}). 
Here we choose to fit these parameters
to reproduce the xc energies of parabolic QDs with $N=6$, which 
is the smallest closed-shell system beyond the simplest $N=2$ case. 
In the external potential $v_{\rm ext}(r)=\omega^{2}r^{2}/2$
we use the confinement strengths $\omega=1/4$ and $1/16$ for which
numerically exact configuration-interaction (CI) data 
is available~\cite{rontani}.
These confinements have a rather wide range with respect to the 
relative weight of the xc effects, and, moreover, the chosen values
are realistic regarding the modeling of real QD devices~\cite{kou,rei}.
The reference xc energy is obtained from
\begin{equation}
E^{\rm ref}_{xc}=E_x^{\rm ref}+E_c^{\rm ref} = E_x^{\rm EXX} + E_{\rm tot}^{\rm exact}-E^{\rm EXX}_{\rm tot}
\label{excref}
\end{equation}
where $E_{\rm tot}^{\rm exact}$ is the reference total energy, $E^{\rm EXX}_{\rm tot}$ is
the total energy from the exact-exchange (EXX) calculation performed here within
the Krieger-Li-Iafrate approximation~\cite{kli} 
and using the {\tt octopus} code~\cite{octopus},
and $E_x^{\rm EXX}$ is the exchange energy. The best fit with 
$E^{\rm ref}_{xc}$ is obtained with parameter values $\gamma=1.12$ and $\alpha=0.45$. 

Next we test if the chosen parameter values yield reasonable and consistent 
results for different 2D systems. This is naturally a desired property in
any density functional in order to be a {\em predictive} approximation.
First we consider parabolic QDs with $N=2\ldots 12$ and $\omega=1/16\ldots 1$.
The results are summarized in Table~\ref{table1}.
\begin{table}
\caption{\label{table1} Exchange-correlation energies for parabolic quantum dots. The optimal parameters
$\gamma=1.12$ and $\alpha=0.45$ have been used for the calculation 
of $E_{\rm xc}^{\rm model}$. The last row contains the mean percentage error.}
\begin{tabular*}{\columnwidth}{@{\extracolsep{\fill}} c c c c c}
\hline
\hline
$N$  & $\omega$ & $-{\rm E}_{\rm xc}^{\rm ref}$  & $-{\rm E}_{\rm xc}^{\rm model}$ & $-{\rm E}_{\rm xc}^{\rm LDA}$ \\
\hline
2   &  1         & $1.246^*$        & 1.195  &  1.174   \\
2   & 1/4        & $0.5987^\dagger$  &  0.5794 &  0.5821  \\
2   & 1/6        & $0.4936^*$        &  0.4678 &  0.4721  \\
2   & 1/16       & $0.2774^\dagger$  &  0.2789 &  0.2820  \\
6   & 1/1.89$^2$ & $2.156^\dagger$   &  2.138  &  2.137   \\
6   & 1/4        & $2.014^\dagger$   &  2.008  &  2.011   \\
6   & 1/16       & $0.9265^\dagger$  &  0.9309 &  0.9429  \\
12  & 1/1.89$^2$ & $4.708^\ddagger$  &  4.716  &  4.701  \\
\hline
$\Delta$  & &    &  $1.86\,\%$   & $2.19\,\%$ \\
\hline
\hline
\end{tabular*}
 \begin{flushleft}
  $^*$ Total energy from the analytic solution in Ref.~\cite{taut}.\\
  $^\dagger$ Total energy from the CI data in Ref.~\cite{rontani}.\\
  $^\ddagger$ Total energy from the QMC data in Ref.~\cite{pederiva}.
  \end{flushleft}
\end{table}
The reference xc energies have been calculated from Eq.~(\ref{excref}) using
the total-energy data from analytic~\cite{taut}, CI~\cite{rontani}, 
and quantum Monte Carlo~\cite{pederiva} (QMC) calculations as indicated in
the table. Overall, we find a very good performance of our functional,
the mean error being $1.86\,\%$, which is smaller than that of
the 2D-LDA ($2.19\,\%$). Although also the LDA is this accurate
for the xc energy, it should be noted that both the exchange and correlation 
parts, respectively, have significant errors (see, e.g., Refs.~\cite{x} 
and~\cite{c}), and the good overall performance follows from the well-known 
error cancellation. It is also noteworthy that the 12-electron case is
very accurate, and it can be expected that the accuracy is preserved
for larger when $N$ is increased further.

\begin{figure}
\includegraphics[width=0.75\columnwidth]{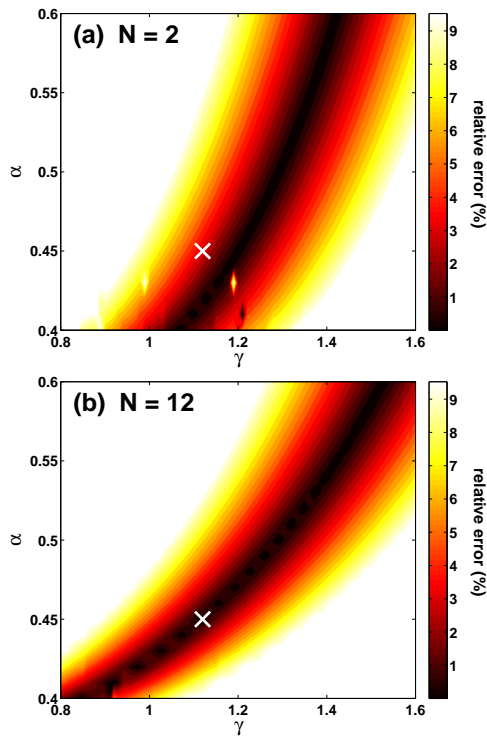}
\caption{(color online). Absolute relative error for parabolic quantum dots, (a) $N=2$, $\omega=1$ and
(b) $N=12$, $\omega=1/1.89^2$ with respect to parameters $\gamma$ and $\alpha$. The crosses mark the values chosen 
from the fit to the $N=6$ case.}
\label{fig1}
\end{figure}

Table~\ref{table1} raises a natural question whether the good performance 
simply follows from the fact that $\gamma$ and $\alpha$
were fitted to a similar system with $N=6$. Therefore, in Figs.~\ref{fig1}(a) 
and (b)
we examine the ``extreme'' cases of Table~\ref{table1} with 
$N=2$ ($\omega=1$) and $N=12$ ($\omega=1.89^2$), respectively.
The figures show the absolute relative errors as functions of both $\gamma$ and $\alpha$, 
so that here the parameter values have been left undetermined for both two cases.
The white crosses show the {\em chosen} values $\gamma=1.12$ and $\alpha=0.45$ 
based on the $N=6$ data.
In both cases, the crosses match very well with the optimal regime where
the relative error is smallest for $N=2$ and $N=12$. 
Hence, Fig.~\ref{fig1} confirms that, at least for parabolic
QDs, the functional is consistent. The figure also demonstrates the
strong correlation between the two parameters as well as the uniqueness between them
-- for each $\gamma$ ($\alpha$) there is only one compatible $\alpha$ ($\gamma$).

\begin{figure}
\includegraphics[width=0.8\columnwidth]{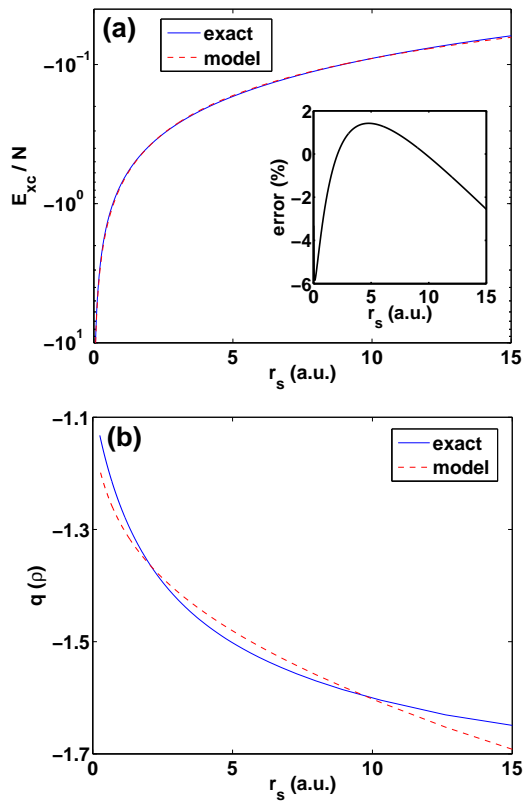}
\caption{(color online). (a) Exchange-correlation energy per particle
for the two-dimensional homogeneous electron gas
obtained using our functional with the chosen parameters $\gamma=1.12$ and $\alpha=0.45$
(dashed line) in comparison with the exact result (solid line). The inset
shows the relative error. (b) Result for the function $q(\rho)$ obtained
using our functional (dashed line) in comparison with the optimal
values required to reproduce the exact exchange-correlation energy of the
two-dimensional homogeneous electron gas.}
\label{fig2}
\end{figure}

Finally we consider the 2DEG corresponding to the important limit of an infinite
electron number. Figure~\ref{fig2}(a)
shows the comparison of the xc energy per particle with respect to 
the exact 2DEG result. Here we have used the same parameter values
$\gamma=1.12$ and $\alpha=0.45$ as before. We find an excellent agreement
through a wide range of the density parameter $r_s=(\pi\rho)^{-1/2}$.
In the realistic density range the relative error is within a few percent 
(see the inset). In Fig.~\ref{fig2}(b) we show the function $q(\rho)$ 
of our functional (dashed line) in comparison with the optimal values
to reproduce the exact xc energy of the 2DEG. Overall, we find good
consistency in the results at varying $r_s$. More importantly, 
regarding the values for $\gamma$ and $\alpha$ the
present functional is also consistent in the comparison between
2DEG and the QDs above.

To summarize, we have used the Colle-Salvetti scheme, and in particular
its recent generalizations to derive an approximate local density functional
for the exchange-correlation energy of electrons in two dimensions.
The functional has a simple polynomial form and it fulfills the
sum-rule constraint of the exchange-correlation hole.
We have fitted the remaining free parameters against exact
results for six-electron quantum dots and found an excellent 
consistency in the results for a set of quantum dots with
varying electron number and varying relative proportion of
the exchange-correlation energy. The functional is precise
also for the two-dimensional homogeneous electron gas with
the same fixed parameters. Therefore, we may expect the functional
to have predictive power in density-functional calculations
for various two-dimensional electron systems. In this respect,
generalization to spin-polarized systems would be the most important 
future extension of the method.

\begin{acknowledgments}
This work was funded by the Academy of Finland.  
\end{acknowledgments}

\end{document}